\documentstyle[11pt,aaspp4,flushrt]{article}
\def \beq{\begin{equation}}
\def \eeq{\end{equation}}
\def \deg{\hbox{$^\circ$}}
\def \lensname{HDFS~2232509--603243}
\begin{document}

\title{A possible gravitational lens in the Hubble Deep Field
        South}
%\footnote{Based on observations made with the NASA/ESA
%        Hubble Space Telescope, which is operated by AURA under NASA
%        contract NAS 5-26555}

\author{Rennan Barkana\footnote{email: barkana@ias.edu},
        Roger Blandford\altaffilmark{4},
        David W. Hogg\footnote{Hubble Fellow}}
\affil{Institute for Advanced Study, Olden Lane, Princeton, NJ 08540, USA}
\altaffiltext{4}{also California Institute of Technology, Pasadena, CA 91125}

\begin{abstract}
We model an apparent gravitational lens system \lensname in the
Hubble Deep Field South. The system consists of a blue $V=25$~mag arc
separated by $0\farcs9$ from a red $V=22$~mag elliptical galaxy.  
A mass distribution which follows the observed light distribution
with a constant mass-to-light ratio can fit the arc component
positions if external shear is added. A good fit is also obtained 
with simple parameterized models, and all the models predict a
fourth image fainter than the detection limit. 
The inferred mass-to-light ratio is $\sim 15$ in solar units if the
lens is at $z=0.6$.  The lens models predict a velocity dispersion
of $\sim 280~{\rm km\ s^{-1}}$ which could be confirmed with 
spectroscopy.
\end{abstract}


\vspace{.1in}

\section{Introduction}
Detailed studies of gravitational lenses have provided a wealth of
information on galaxies, both through modeling of individual
lens systems (e.g., Schneider, Ehlers, \& Falco 1992; 
Blandford \& Narayan 1992) and from the statistical properties of 
multiply imaged sources (e.g., Turner, Ostriker, \& Gott 1984; 
Kochanek 1996).
The extraordinary resolution of the Hubble Space Telescope (HST) has
allowed for the first time the detection of lenses where both the
source and the deflector are ``normal'' optically-selected galaxies.
Ratnatunga et al.\ (1995) discovered two 
four-image ``Einstein cross'' gravitational lenses in the
Groth-Westphal strip. Their mass models required ellipticities 
substantially above those
of the light distributions, suggesting (as is commonly suspected in
radio lenses) that some external shear is important to the lensing.

The Hubble Deep Field (HDF, Williams et al.\ 1996), the deepest set of
exposures taken with the HST, has enabled an unprecedented study of
galaxy morphological evolution (e.g., Ellis 1998) and the global star
formation history (e.g., Madau, Pozzeti, 
\& Dickinson 1998). Candidate gravitational
lenses were reported by Hogg et al.\ (1996), who noted on the basis of
radio surveys that several lenses are expected in the HDF
(cf.\ Cooray, Quashnock, \& Miller 1998). However,
after further observations not one strong case for multiple imaging
has been identified (Zepf, Moustakas, \& Davis 1997; Cohen 1998).

The Southern Hubble Deep Field (HDF-S, Williams et al.\ 1998) is a
second set of ultra-deep HST exposures, and several lenses are again 
expected. The best candidate lens,
noted by the HDF-S team (Williams 1998), consists of
a blue arc about $0\farcs9$ from an elliptical galaxy. In this 
{\sl Letter} we report on a detailed study of
this arc and argue that it is a bona fide multiply imaged source.

\section{Observations}

The WFPC2 HDF-S is a $\sim 5$~arcmin$^2$ field centered on
$22\,32\,56.22~-60\,33\,02.69$ (J2000), imaged with the WFPC2
instrument on the HST in the $F300W$, $F450W$, $F606W$, and 
$F814W$ bandpasses
for total exposure times of $1.4\times 10^5$, $1.0\times 10^5$,
$8\times 10^4$, and $1.0\times 10^5$~s respectively, reaching
point-source sensitivities of roughly $V=29$~mag (Williams et al.\ 
1998). For the ``version-1.0'' reductions of the images, 
we adopt the Vega-relative calibrations supplied by the HDF-S team
whereby 1~DN corresponds to 
$F300W=19.43$~mag, $F450W=22.02$~mag, $F606W=22.90$~mag, or
$F814W=21.66$~mag. The images clearly show a
bright candidate lens system at pixel location $(2982.7,2837.6)$.
With the pixel scale of $0\farcs0398$ this is also 
$22\,32\,50.90~-60\,32\,43.0$ (J2000).  This candidate lens system
will be called ``\lensname'' hereafter.

\begin{figure}
\epsscale{0.8}
%\plotone{hdfslens.ps}
\plotone{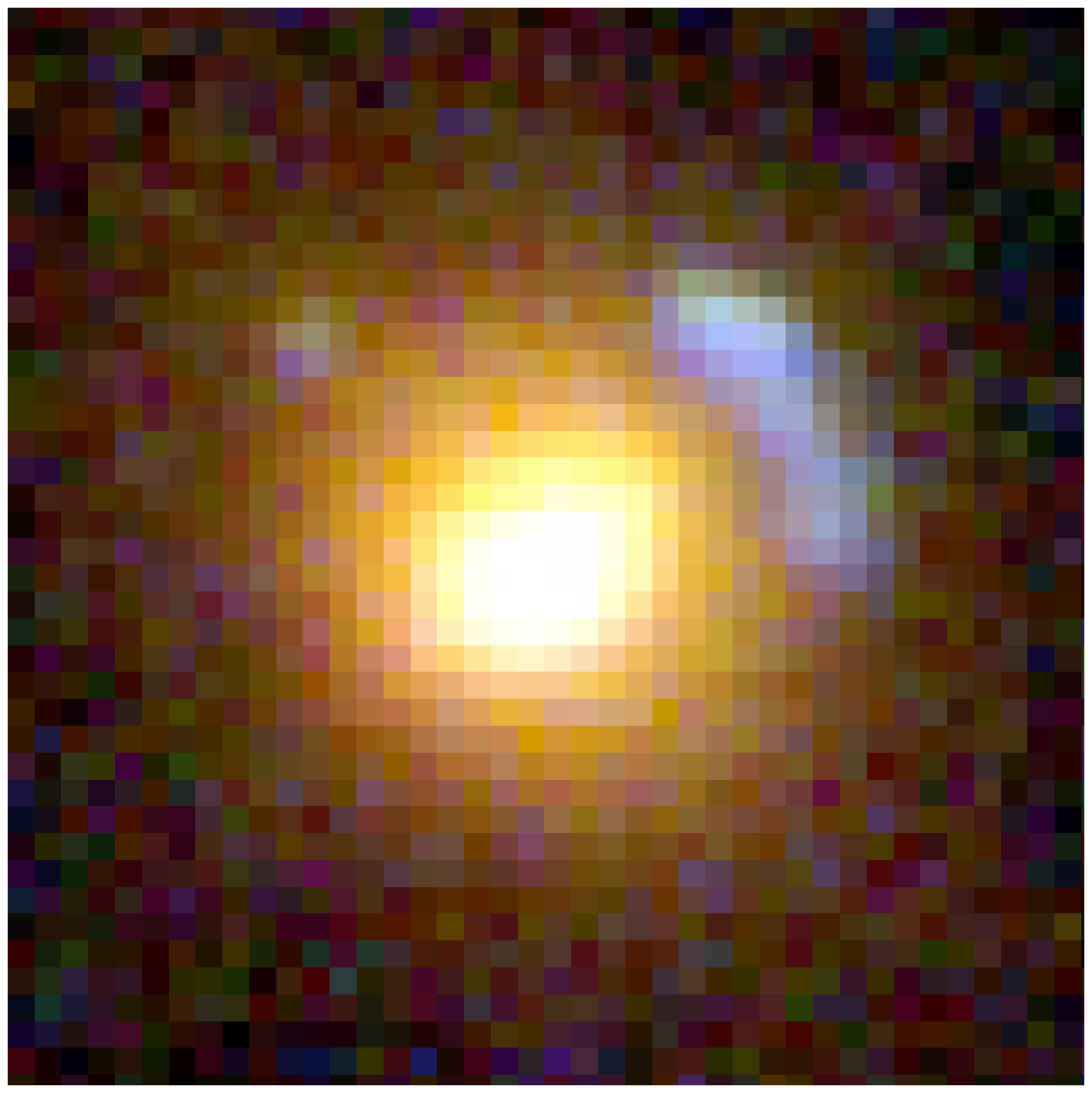}
\caption{A region taken from the public release of the HDF-S 
(Williams et al.\ 1998). It shows the lens galaxy, the arc to its
NW, and the dot to its NE. North is up and East is to the left
(except for a $0.5\deg$ rotation). The figure is centered on
the lens galaxy and measures $3\farcs2$ on a side.}
\label{fig:lens}
\end{figure}

The morphology of \lensname is that of a $V\sim 22$~mag
elliptical galaxy (hereafter ``the elliptical'') with a much bluer
arc (hereafter ``the arc'') located $0\farcs9$ to the NW from the 
center of the elliptical, as shown in Figure~\ref{fig:lens}. 
There is also a faint
point-like source (hereafter ``the dot'') located $0\farcs96$
to the NE from the center of the elliptical.

The elliptical is smooth and symmetrical, so images created by
rotating its images through $180\deg$ (centered on the pixel
nearest the center of the elliptical) and subtracting them from their
unrotated originals show essentially no sign of the elliptical.
Photometry of the arc was
performed in these ``rotated-subtracted'' images.  The total $F606W$
flux (given in Table~\ref{tab:photometry})
was measured through a partial annulus of inner radius
$0\farcs517$ and outer radius $1\farcs194$, extending over position
angles $0\deg$ (north through east) through $-108\deg$.  Colors 
were measured through a partial
annulus of inner and outer radii $0\farcs677$ and $1\farcs035$, 
over $-18\deg$ through $-90\deg$.  The flux and colors 
of the dot were measured through a $0\farcs159$
radius circular aperture. Because nearby pixels are correlated,
signal-to-noise ratios for photometric measurements 
were found by re-binning the image into pixels $4\times 4$ times 
larger.

The arc separates into four distinct components, labeled $A$,
$B$, $C$ and $D$, going clockwise around the arc.  All 
have very similar colors.  Components $A$, $B$, and $D$ appear to have
point-like centers with fuzz, while component $C$ is more extended.
Gaussians with the same FWHM as the PSF (4~pixels) were fit to
components $A$, $B$ and $D$, and a gaussian of variable width was fit to
component $C$, all in the $F450W$ rotated-subtracted image.  The best-fit
positions and fluxes of these gaussians are given in
Table~\ref{tab:astrometry}. 

In order to make arc-free images of the elliptical, the region of each
image in the partial annulus centered on the elliptical
with inner and outer radii of $0\farcs517$ and $1\farcs194$ and position
angles $0\deg$ through $-108\deg$ was replaced with the corresponding
region diametrically opposite (i.e., at a position angle rotated by
$180\deg$).  These ``arc-removed'' images show virtually no evidence of
the arc.  The total $F814W$ flux of the elliptical was measured out to
its 2-$\sigma$ (per pixel) isophote.  The colors were measured through
$0\farcs915$ radius circular apertures, and are given in
Table~\ref{tab:photometry}.

A de~Vaucouleurs profile is a very good fit to the azimuthally 
averaged radial profile of the elliptical in the ``arc-removed'' 
images. The effective radius is $r_e= 0\farcs31$ and the 
$F814W$-band surface brightness
at the effective radius is $21.0$~mag in each arcsec$^2$.  The
isophote ellipticity orientation varies with radius.  The outer
isophotes (radii $>0\farcs8$) of the elliptical have ellipticity $\sim
0.2$ and are elongated along $+40\deg$.  The inner isophotes (radii
$<0\farcs3$) also have ellipticity $\sim 0.2$ but are elongated along
$-40\deg$, i.e., almost orthogonal. Isophotes intermediate to these
two regimes are fairly circular.

Comparison of the colors of the elliptical with HDF galaxies with
known redshifts suggests that its redshift is roughly $z=0.6$
(cf.\ the photometric redshift of 0.5 of Gwyn (1998)).
This suggests a total luminosity\footnote{We assume a world 
model of $\Omega=0.3$ in matter and no cosmological constant,
and also set $H_0 = 100\, h\, {\rm km\ s^{-1}\,Mpc^{-1}}$.}
of $8.7\times 10^9 h^{-2} L_{\sun}$ in the rest-frame B-band, 
or roughly $L_{\ast}$. The luminosity is a factor of 3 larger
if $z=0.8$, and a factor of 6.5 smaller if $z=0.3$.

\section{Lens models}

We consider several models for the lensing mass distribution,
simple parameterized models as well as a constant mass-to-light
ratio model. 
The image positions are not
very sensitive to the radial profile of the lens, so for the
parameterized models we take it
to be singular and isothermal. Asymmetry is required 
and we consider two types, external shear or 
an elliptical galaxy.
We assume that the arc components $A$, $B$ and $D$ are three
images of a common source, with $A$ and $B$ a merging pair. 
Component $C$ can easily be produced
if the source is extended or multi-component. We do not include
as constraints component $C$ or the observed flux ratios among
components $A$, $B$ and $D$, since these are sensitive to the detailed
flux distribution of the source. We discuss the dot below.

We consider a lens at redshift $z_L$ and a source at $z_S$. Then 
the lens mass distribution determines a convergence $\kappa$
and a potential $\psi$
(e.g., Schneider et al.\ 1992; Blandford \& Narayan 1992).
The singular isothermal sphere (SIS) is defined by
$\kappa=b/[2 r]$. We add to the potential the
term $\psi_{\gamma}=- \frac{1}{2}\gamma r^2 \cos 2 (\theta-
   \theta_{\gamma})\ , $
which defines an external shear of magnitude $\gamma$ and 
direction $\theta_{\gamma}$. If the shear is
due to an axisymmetric galaxy then it is located
at an angle $\theta_{\gamma}$ from the 
lens galaxy. We denote by SIS$+\gamma$ the 
SIS model with external shear.

A singular isothermal elliptical mass distribution (SIEMD) 
with an axis ratio $a$ and a major axis along the $y$-axis
is given by $\kappa=b/[2 \sqrt{x^2+y^2 a^2}]$. 
More generally we rotate the major axis by an angle 
$\theta_{\epsilon}$. The lensing properties 
of the SIEMD have been studied by Kassiola \& Kovner
(1993) and by Kormann, Schneider, \& Bartelmann (1994). 

We fit these two lens models, the SIS$+\gamma$ model and
the SIEMD, to the three observed image positions, with
results shown in Table~\ref{tab:models}. 
Both lens models have an ellongated diamond caustic with
four cusps. We indicate a source near a cusp on the major 
axis by [maj] and one near a minor cusp by [min].
Since the observed position $B$ 
is farther from the center of the lens than $A$ or $D$, the
arc is more naturally produced by a minor cusp for which
only the middle image of the three is outside the critical
curve. 

\begin{figure}
\epsscale{0.8}
%\plotone{caus8col.ps}
%\plotone{caus8col.ps}
\plotone{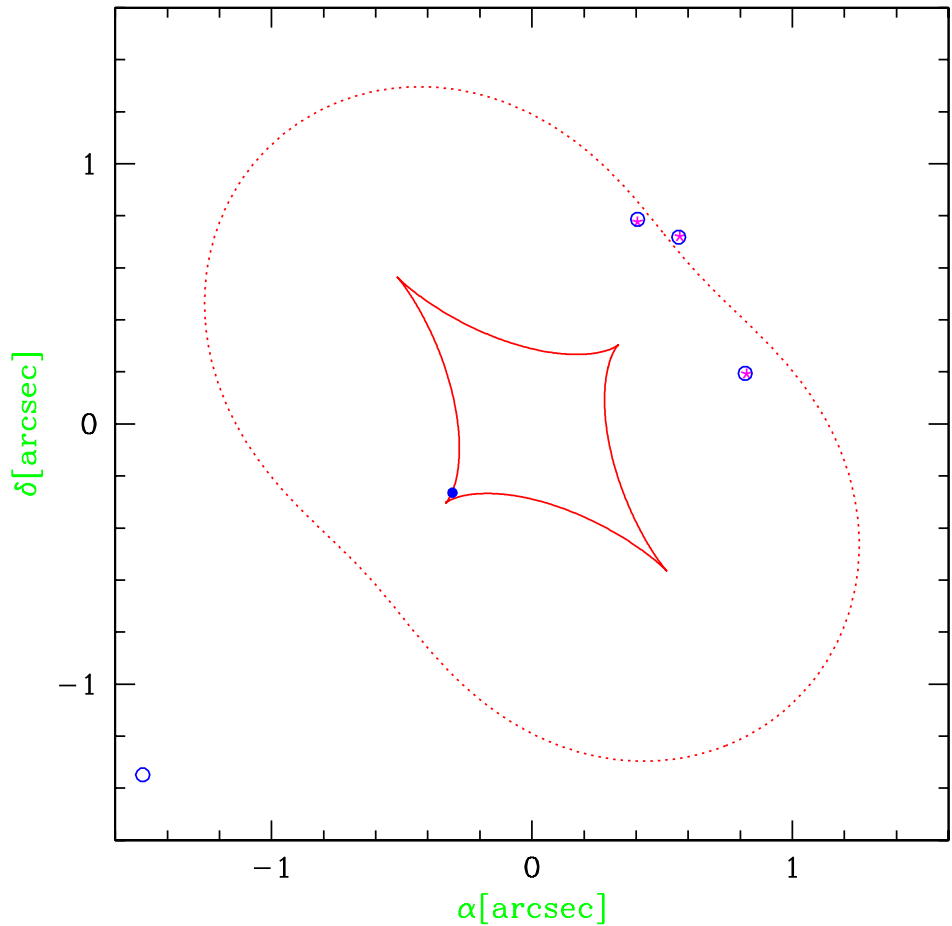}
\caption{Image and source positions for the SIS$+\gamma$[min]
model (column 3 of Table~\ref{tab:models}). Four image 
positions ($\circ$) are indicated along with the source
position ($\bullet$) and the observed positions of 
arc components $A$, $B$ and $D$
($\star$). Also shown are the tangential 
caustic (solid) and critical curve (dotted). 
The figure is centered on the lens galaxy, with the 
$\alpha$-direction west and the $\delta$-direction north.}
\label{fig:gamma}
\end{figure}

Interestingly, the orientation of 
the SIEMD[min] model agrees with the observed
orientation of the elliptical's outer isophotes, while
the orientation of the SIEMD[maj] model agrees with 
that of the inner isophotes. However, 
among the simple parameterized models only the SIS$+\gamma$[min]
model fits the data well (see Table~\ref{tab:models} and
Figure~\ref{fig:gamma}).
This model requires a substantial shear 
$(\gamma=0.26)$ produced by objects lying in the
NE or SW directions. To compute the mass-to-light
ratio for various models, we fix a circle about the galaxy 
center of radius $0\farcs915$. The luminosity of the galaxy 
within this circle is $7.6\times 10^9 h^{-2} L_{\sun}$ if 
$z_L=0.6$. If, e.g., $z_S=1.5$, then 
the mass within the same circle of the SIS is 
$M=2.1 \times 10^{11} h^{-1} (b/1\arcsec) M_{\sun}$. This
yields a mass-to-light ratio of $27\, h\, (b/1\arcsec)$,
and a velocity dispersion $\sigma=270 \sqrt{b/1\arcsec}\  
{\rm km\ s^{-1}}$. 

%Note that for the lenses which they 
%considered, Keeton, Kochanek, \& Seljak (1997)
%found a substantial improvement in the fits with a model 
%which includes ellipticity as well as external shear, but the
%data on \lensname\ cannot yet constrain such a complex
%lens model. 

\begin{figure}
\epsscale{0.8}
%\plotone{caus11col.ps}
%\plotone{caus11.ps}
\plotone{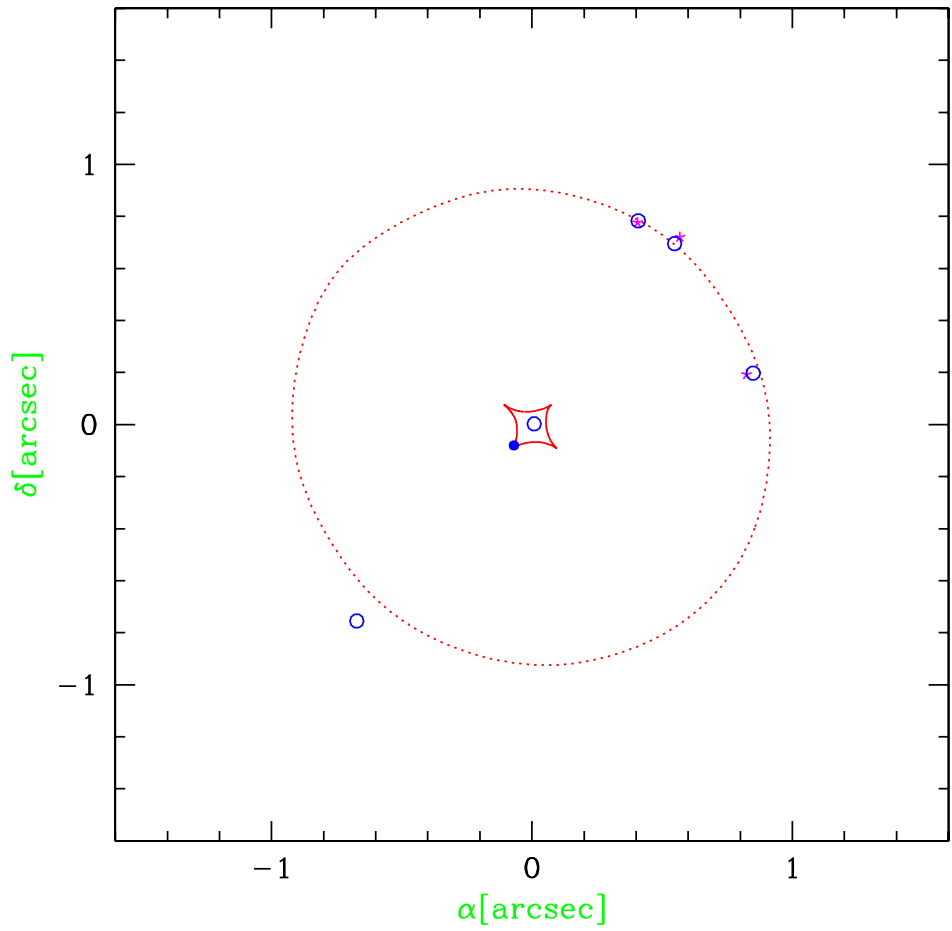}
\caption{Same as Figure 2, but for the M/L$+\gamma$ model
with $\gamma=0.05$ (last column of 
Table~\ref{tab:models}). There are five images, although
the one near the lens center is extremely faint.}
\label{fig:MtoL}
\end{figure}

As a second approach to modeling, we utilize the resolved
image of the lens galaxy (smoothed with a 4-pixel FWHM 
gaussian) to construct a model with
a constant mass-to-light ratio (denoted M/L). The pure M/L
model fails but together
with external shear it produces an excellent fit,
albeit with a large shear. Even a small shear
such as $0.05$ produces a substantially better fit than 
with no shear (see Table~\ref{tab:models} 
and Figure~\ref{fig:MtoL}).
All the M/L models produce a minor cusp configuration. 
The parameter $\bar{\kappa}$ for the M/L models gives
the mean $\kappa$ in a circle of radius $0\farcs915$ about
the galaxy center. If $z_L=0.6$ and $z_S=1.5$ then 
the mass within this circle is 
$M=1.9 \times 10^{11} h^{-1}\, \bar{\kappa}\, M_{\sun}$, 
which yields a mass-to-light ratio of $25\, h\, \bar{\kappa}$.
If we assume a small shear then $\bar{\kappa}\sim 1$
(Table~\ref{tab:models}), and the mass-to-light ratio is
on the high side of the range of other lens galaxies 
(Keeton, Kochanek, \& Falco 1998). Note, however, that we only 
have an estimated photometric redshift for the lens and
a guess for the source redshift. The mass-to-light ratio
changes to $14\, h\, \bar{\kappa}$ for $(z_L=0.8,\ z_S=2.0)$
and to $57\, h\, \bar{\kappa}$ for $(z_L=0.3,\ z_S=1.0)$.

No fourth image is detected down to the detection threshold 
of about 28.9 mag in $F606W$. This is consistent with the
models, which predict a fourth image fainter than the third
by a factor of 3 or 4. The M/L models are non-singular and 
predict a fifth image fainter
by an additional factor of $\sim 100$. The best-fitting
models also locate the image of the cusp near the observed
position of component $C$, so this component can be produced
if the source is extended.

Thus far we have not included the dot, assuming it to be a
separate background source. 
We consider also the possibility that the dot is the third image 
corresponding to components $A$ and $B$. The types of lens models 
used above can fit these alternative image 
positions with similar $\chi^2$ values as in Table 3,
but they fail since they predict an observable arc 
which includes counter images of components $C$ and $D$ 
and extends South-East from the dot. More elaborate
lens models could work but only if the sources of 
components $C$ and $D$ are small (subpixel) and lie right 
on the caustic.

\section{Discussion}

We have modeled a candidate gravitational lens, \lensname, found in 
the HDF-S. The image positions can be fit in a minor cusp
configuration by simple lens models
which include ellipticity or shear, or with shear added to a
model with a constant mass-to-light ratio. 
However, an exact fit is obtained
only with a substantial shear produced by objects lying to the
NE or SW. No obvious candidate galaxy lies within $5\arcsec$ of 
the lens, but there are several galaxies
$\sim 20\arcsec$ to the SW and others $\sim 35\arcsec$
to the NE with photometric redshifts in the range 0.5--0.7
(Gwyn 1998). Spectroscopic redshifts will determine the
possible presence of galaxy groups. 

The lensing galaxy is expected to have a redshift $z\sim 0.6$,
which gives it a luminosity of $\sim L^{\ast}$. Lens models 
imply a mass-to-light ratio of $\sim 25\ h$, and a velocity 
dispersion of $\sim 280~{\rm km\ s^{-1}}$. If the elliptical falls 
on the fundamental plane, a central
line-of-sight velocity dispersion of only $\sim 100~{\rm km\ s^{-1}}$
is expected if $z=0.6$ (Jorgensen, Franx \& Kjaergaard 1996),
which suggests that $z\sim 0.8$ is more likely.

Showing that the arc and the lens galaxy are at different
redshifts would rule out the possibility that the arc
is a tidally disrupted satelite.
On 10-m--class telescopes, it is straightforward to get spectroscopic
redshifts of galaxies at $22$~mag, even elliptical galaxies which tend
to have absorption-feature dominated spectra (Cohen et al.\ 1998).  Some
elliptical galaxies at this flux level even have measured central
velocity dispersions (van~Dokkum et al.\ 1998; Pahre 1998).
Thus the prospects for follow-up
spectroscopy on the elliptical are good.  It is common to obtain
redshifts for sources at $25$~mag, even some which are close to bright
nearby sources (Steidel et al.\ 1996). A redshift for the arc is
likely obtainable if it is at $z<1.3$ since 
its blue color suggests that it will show 
strong emission lines in a visual spectrum (Hogg et al.\ 1998). 
The arc is unlikely to be at redshift $z>2.3$ since
it shows significant $F300W$ flux.

%The fact that \lensname\ appears to be comprised of a background
%galaxy lensed by a field elliptical is notable given the paucity of
%lenses in the HDF (see \S 1) and the smaller number of ellipticals in
%the HDF-S compared to the HDF (Williams 1998).

\acknowledgements It is a pleasure to thank Bob Williams and the
entire HDF-S team for taking, reducing, and making public the
beautiful images of the HDF-S.  Scott Burles, Wayne Hu and 
Inger Joergensen provided very useful help in short order.  
We thank the referee Emilio Falco for useful comments.
This research is based on
observations made with the NASA/ESA Hubble Space Telescope, which is
operated by AURA under NASA contract NAS 5-26555.  Barkana and
Blandford acknowledge
support by Institute Funds.  Hogg acknowledges Hubble Fellowship grant
HF-01093.01-97A from STScI, which is operated by AURA under NASA
contract NAS~5-26555.

%%%% FIGURES
\vfill\eject

%%%%%%%%Figure 1
%%%%%%%%

%%%%%%%%Figure 2
%%%%%%%%

%%%%%%%%Figure 3
%%%%%%%%

%%%% TABLES
\vfill\eject

\begin{deluxetable}{lrlrlrlrl}
\tablecaption{Photometry}
\tablehead{
  \colhead{component} &
  \multicolumn{2}{c}{flux\tablenotemark{a}} &
  \multicolumn{2}{c}{$F300W-F450W$} &
  \multicolumn{2}{c}{$F450W-F606W$} &
  \multicolumn{2}{c}{$F606W-F814W$} \\
  &
  \colhead{(mag)} &
  \colhead{[snr\tablenotemark{~b}]} &
  \colhead{(mag)} &
  \colhead{[snr1,snr2\tablenotemark{~c}]} &
  \colhead{(mag)} &
  \colhead{[snr1,snr2]} &
  \colhead{(mag)} &
  \colhead{[snr1,snr2]}
}
\startdata
elliptical & $F814W= 20.52$ & [950] &
    0.31  & [9.2,99] &
    1.86  & [99,610] &
    1.53  & [610,1200] \nl
arc        & $F606W= 25.19$ & [41] &
  $-0.96$ & [11,37] &
    0.24  & [37,51] &
    0.40  & [51,37] \nl
dot        & $F606W= 28.70$ & [6.2] &
  \nodata & [$-1.5$,3.9] &
    0.39  & [3.9,6.2] &
  $-0.05$ & [6.2,2.9]
\enddata
\tablenotetext{a}{Different components have flux measurements in
different bands and through different focal-plane apertures; see text
for details.}
\tablenotetext{b}{Signal-to-noise ratios (snr) measure the flux
through the aperture relative to the expected sky noise rms through
that same aperture.  See text for aperture definitions.}
\tablenotetext{c}{The snr values snr1 and snr2 are for the two
bandpasses involved in the color measurement, bluer first.}
\label{tab:photometry}
\end{deluxetable}

\begin{deluxetable}{lccc}
\tablecaption{Astrometry\tablenotemark{a} and Photometry}
\tablehead{
  \colhead{sub-component} &
  \colhead{$\Delta\alpha$} &
  \colhead{$\Delta\delta$} &
  \colhead{$F450W$\tablenotemark{b}} \nl
  &
  \colhead{(arcsec)} &
  \colhead{(arcsec)} &
  \colhead{(mag)}
}
\startdata
%
% This is rotated from the pixels by 0.5 a degree:
%
elliptical & 0 & 0 & \nodata \nl
arc-$A$      & $+0.406$ & $+0.777$ & 27.55 \nl
arc-$B$      & $+0.568$ & $+0.721$ & 27.43 \nl
arc-$C$      & $+0.711$ & $+0.478$ & 27.08 \nl
arc-$D$      & $+0.824$ & $+0.192$ & 28.10 \nl
dot        & $-0.778$ & $+0.567$ & \nodata
% Unrotated dot: -.773, .574
\enddata
\tablenotetext{a}{Positions are given relative to $22\,32\,50.9017$
$-60\,32\,43.009$ (J2000), with the $\alpha$-direction west and the
$\delta$-direction north.  We estimate position errors of $0\farcs02$
from fitting an artificial image.}
\tablenotetext{b}{The magnitudes of the arc components are
based on the fluxes of the fitted gaussians.  The fluxes do not add up
to the total arc flux because the magnitudes exclude some diffuse
emission.}
\label{tab:astrometry}
\end{deluxetable}

\begin{deluxetable}{lccccccc}
\tablecaption{Best fit model results}
\tablehead{ \colhead{Parameter} & \colhead{SIEMD[min]} 
& \colhead{SIEMD[maj]} & \colhead{SIS$+\gamma$[min]}
& \colhead{SIS$+\gamma$[maj]} & \colhead{M/L} & 
\colhead{M/L$+\gamma$} & \colhead{M/L$+\gamma$}}
\startdata
$b$ or $\bar{\kappa}$ & $0\farcs85$ & $0\farcs84$ & $1\farcs08$ & 
$0\farcs87$ & 1.10 & 1.40  & 0.998 \nl
$a$ or $\gamma$ & $0.03_{-.03}^{+.97}$ & $0.96_{-.74}^{+.04}$ & 
$0.26_{-.05}^{+.04}$ & $0.005_{-.005}^{+.18}$ & \nodata &
$0.52_{-.14}^{+.15}$ & 0.05 \nl
$\theta_{\epsilon}$ or $\theta_{\gamma}$ & $41\fdg 2_{-2.6}^{+2.9}$ & 
$-48\fdg6_{-3.1}^{+3.0}$ & $42\fdg5_{-.7}^{+.7}$ & $-49\fdg0_{-2.8}^
{+3.5}$ & \nodata & $43\fdg5_{-.5}^{+.5}$ & $53\fdg4$ \nl
\#dof & 1 & 1 & 1 & 1 & 3 & 1 & 1 \nl
$\bar{\chi}^2$ & 5.1 & 5.6 & 0.26 & 5.8 & 97 & 0.40 & 4.5 \nl
\enddata       
\tablecomments{The best fits are formally at $a \rightarrow 0$
for the SIEMD[min] model, at $a \rightarrow 1$ for the
SIEMD[maj] model, and at $\gamma \rightarrow 0$ for the
SIS$+\gamma$[maj] model. The overall best fit for the M/L$+\gamma$ 
model is given in column 6, while column 7 gives the best fit for this
model with the restriction $\gamma=0.05$. Where given, parameter 
ranges indicate $1\sigma$ uncertainties defined by $\Delta \chi^2=
\bar{\chi}^2$. All angles are measured north through east.}
\label{tab:models}
\end{deluxetable}

\end{document}